\input amstex.tex
\documentstyle{amsppt}
\def\wt{\operatorname{wt}}
\def\Tr{\operatorname{Tr}}
\newcount\refcount
\advance\refcount 1
\def\newref#1{\xdef#1{\the\refcount}\advance\refcount 1}
\newref\crssII
\newref\conwaysloane
\newref\knilllaflamme
\newref\macwilliamssloane
\newref\unitaryenum
\newref\shadowboundcodes
\newref\shorlaflamme
\topmatter
\title Quantum shadow enumerators \endtitle
\author Eric Rains\endauthor
\affil AT\&T Research \endaffil
\address AT\&T Research, Room 2D-147, 600 Mountain Ave.
         Murray Hill, NJ 07974, USA \endaddress
\email rains\@research.att.com \endemail
\date October 30, 1996 \enddate
\abstract
In a recent paper [\shorlaflamme], Shor and Laflamme
define two ``weight enumerators'' for quantum error correcting codes,
connected by a MacWilliams transform, and use them to give a
linear-programming bound for quantum codes.  We extend their work by
introducing another enumerator, based on the classical theory of shadow
codes, that tightens their bounds significantly.  In particular, nearly all
of the codes known to be optimal among additive quantum codes (codes
derived from orthogonal geometry ([\crssII])) can be shown to be optimal
among all quantum codes.  We also use the shadow machinery to extend a
bound on additive codes ([\shadowboundcodes]) to general codes,
obtaining as a consequence that any code of length $n$ can
correct at most $\lfloor {n+1\over 6}\rfloor$ errors.
\endabstract
\endtopmatter
\head Introduction\endhead

One of the basic problems in the theory of quantum error correcting codes
(henceforth abbreviated QECCs) is that of giving good upper bounds on the
minimum distance of a QECC.  The strongest technique to date for this
problem is the linear programming bound introduced by Shor and Laflamme
(\cite\shorlaflamme).  Their bound involves the definition of two ``weight
enumerators'' for a QECC; the two enumerators satisfy certain inequalities
(e.g., nonnegative coefficients), and are related by MacWilliams
identities.  This allows linear programming to be applied, just as for
classical error correcting codes (\cite\macwilliamssloane).

Linear programming was first applied to bounds for quantum codes in
\cite\crssII, which gave bounds only for codes of the type introduced in that
paper (henceforth denoted ``additive'' codes).  The linear programming
bound given there essentially consists of three families of inequalities.
Two of these were generalized to arbitrary quantum codes in
\cite\shorlaflamme; the current paper generalizes the third.  Consequently, in
the table of upper bounds given in \cite\crssII, all but 10 apply in general;
it follows that nearly all of the codes known to be optimal among additive
codes are optimal among QECCs in general.

A quick note on terminology: We will be using the terms ``pure'' and
``impure'' in place of the somewhat cumbersome terms ``nondegenerate'' and
``degenerate''; that is, a pure code is one in which all low weight
errors act nontrivially on the codewords.

\head 1. Quantum weight enumerators\endhead

Recall that a quantum code $\Cal{C}$ is a $K$-dimensional subspace of
a $2^n$-dimensional Hilbert space $V$; $\Cal{C}$ has minimum distance $d$
if and only if
$$
\langle v|U_{d-1}|v\rangle=\langle w|U_{d-1}|w\rangle,
$$
for $v$ and $w$ ranging over all unit vectors in $\Cal{C}$
(\cite\knilllaflamme), and for $U_{d-1}$ ranging over all $d-1$ qubit errors.
We will use the notation $((n,K,d))$ to refer to such a code.  We will
follow the convention of \cite\crssII, in that a $((n,1,d))$ must be
pure.

To verify that a code has minimum distance $d$, it suffices to restrict
one's attention to errors of the form
$$
\sigma_1\otimes \sigma_2\otimes \cdots \otimes \sigma_n,
$$
where each $\sigma_i$ ranges over the set
$$
\left\{ \pmatrix 1&0\\ 0&1\endpmatrix\!,
        \sigma_x=\pmatrix 0&1\\ 1&0\endpmatrix\!,
        \sigma_y=\pmatrix 0&-i\\ i&0\endpmatrix\!,
        \sigma_z=\pmatrix 1&0\\ 0&-1\endpmatrix \right\};
$$
we will denote the set of such errors by $\Cal{E}$.  For an error $E$ in
$\Cal{E}$, we define the weight $\wt(E)$ of $E$ as the number of the
$\sigma_i$ not equal to the identity.  Also, as in \cite\crssII, we note
that $\Cal{E}$ has the structure of a vector space ${\Bbb F}_2^{2n}$,
with a symplectic bilinear form given by
$$
(-1)^{\langle E_1,E_2\rangle}=E_1 E_2 E_1 E_2.
$$

The weight enumerators of Shor and Laflamme can be defined as follows:
Let $M_1$ and $M_2$ be Hermitian operators on the state space $V$. Then
define
$$
\eqalign{
A_d(M_1,M_2)&=\sum_{E\in \Cal{E}\atop \wt(E)=d} \Tr(E M_1)\Tr(E M_2)\cr
B_d(M_1,M_2)&=\sum_{E\in \Cal{E}\atop \wt(E)=d} \Tr(E M_1 E M_2).\cr}
$$
Note that this differs from the definition in \cite\shorlaflamme\ by
normalization factors, in order to simplify the theory.  After
Shor and Laflamme, we also define two polynomials $A(x,y)$ and
$B(x,y)$ by
$$
\eqalign{
A(x,y)&=\sum_{0\le d\le n} A_d(M_1,M_2) x^{n-d} y^d\cr
B(x,y)&=\sum_{0\le d\le n} B_d(M_1,M_2) x^{n-d} y^d.}
$$

We have the following theorems, from \cite\shorlaflamme:

\proclaim{Theorem 1 (Duality)}
Let $M_1$ and $M_2$ be any Hermitian operators on $V$.
Then
$$
\eqalign{
B(x,y)&=A({x+3y\over 2},{x-y\over 2})\cr
A(x,y)&=B({x+3y\over 2},{x-y\over 2}).\cr}
$$
\endproclaim

\proclaim{Theorem 2 (Bounds)}
Let $P$ be the orthogonal projection onto a $((n,K,d))$.
Then
$$
\eqalign{
&A_0(P)=K,\cr
&A_i(P)\ge 0,\ (0\le i\le n)\cr
&B_0(P)=K^2,\cr
&K B_i(P)-A_i(P)=0,\ (0\le i<d)\cr
&K B_i(P)-A_i(P)\ge 0,\ (d\le i\le n).\cr}
$$
\endproclaim

We will also need the following result:

\proclaim{Theorem 3}  Let $M_1$ and $M_2$ be any positive semi-definite
Hermitian operators on $V$.  Then $B_d(M_1,M_2)\ge 0$ for $0\le d\le n$.
\endproclaim

\demo{Proof}
$B_d(M_1,M_2)$ is a sum of terms of the form
$$
\Tr(M_1 E M_2 E^{-1}).
$$
Each of these terms is the trace of the product of two positive
semi-definite Hermitian operators, and is thus nonnegative.
\qed\enddemo

\head 2. Additive codes \endhead

Before presenting the shadow enumerator, it is instructive to examine a
special case, namely that of additive codes (\cite\crssII).  An additive code
$\Cal{C}$ is derived from a subspace $C$ of $GF(2)^{2n}$, weakly self-dual
under the symplectic inner product; the orthogonal projection onto
$\Cal{C}$ is then of the form
$$
P=2^{-\dim(C)} \sum_{E \in C} s(E) E,
$$
where $s(E)$ are appropriately chosen signs (in particular, $s(1)=1$).

For additive codes, $A_d$ and $B_d$ have combinatorial interpretations.
Indeed,
$$
\Tr(E P)\Tr(E P)=\cases 2^{2(n-\dim(C))}&\text{$E\in C$}\\
                        0&\text{otherwise}\endcases
$$
and
$$
\eqalign{
\Tr(E P E P)&=2^{-2\dim(C)} \sum_{E'\in C} Tr(E E' E E')\cr
            &=2^{n-2\dim(C)} \sum_{E'\in C} (-1)^{\langle E,E'\rangle}\cr
            &=\cases 2^{n-\dim(C)} & \text{$E\in C^\perp$}\\
                     0&\text{otherwise}\endcases\cr}
$$
Consequently, $2^{-2(n-\dim(C))}A_d$ counts the number of elements of
$C$ of weight $d$, while $2^{-(n-\dim(C))}B_d$ counts the number of
elements of $C^\perp$ of weight $d$.

There is a third combinatorial object that we can count, namely the
``shadow'' $S(C)$ of $C$.  This is defined as the set of all $E\in \Cal{E}$
such that
$$
\langle E,E'\rangle\equiv \wt(E')\pmod 2
$$
for all $E'\in C$.  This is completely analogous to the definition of the
shadow of a classical binary code (\cite\conwaysloane).  The primary relevance
of the shadow is that its enumerator can be computed from the
ordinary enumerator:

\proclaim{Theorem 4}
Let $S_d$ be $2^{n-\dim(C)}$ times the number of elements of $S(C)$ of
weight $d$, and define
$$
S(x,y)=\sum_{0\le d\le n} S_d x^{n-d} y^d.
$$
Then
$$
S(x,y)=A({x+3y\over 2},{y-x\over 2}).
$$
\endproclaim

\demo{Proof}
Let us distinguish two cases.  Either $C$ contains an element of odd
weight, or it does not.  In the latter case, an error $E$ is in 
$S(C)$ if and only if it is in $C^\perp$; moreover, $A(x,y)=A(x,-y)$.
So
$$
S(x,y)=B(x,y)=A({x+3y\over 2},{x-y\over 2})=A({x+3y\over 2},{y-x\over 2}).
$$

Thus, assume $C$ contains an element of odd weight.  Since $C$ is weakly
self-dual, it follows that the subset $C_0$ of $C$ consisting of
elements of even weight is, in fact, a subspace of codimension 1;
let it have weight enumerators $A^{(0)}$ and $B^{(0)}$.
Then $S(C)$ can be written as $C^\perp_0-C^\perp$.  In terms of
the weight enumerators, we have:
$$
\eqalign{
2^{n-\dim(C)} S(x,y)&=2^{n-\dim(C_0)} B^{(0)}(x,y)-2^{n-\dim(C)} B(x,y)\cr
                    &=2^{n-\dim(C_0)} A^{(0)}({x+3y\over 2},{x-y\over 2})-
                      2^{n-\dim(C)} A({x+3y\over 2},{x-y\over 2}).\cr}
$$
But $2 A^{(0)}(x,y)=A(x,y)+A(x,-y)$, so
$$
S(x,y)=A({x+3y\over 2},{y-x\over 2}).
$$
\qed\enddemo

Before we proceed to general codes, it will be helpful to digress
momentarily, and consider the following problem: When is an additive code
real?  More generally, when is it {\it equivalent} to a real code?

To answer the first question, recall that
$$
P=2^{-\dim(C)} \sum_{E \in C} s(E) E,
$$
Thus
$$
\overline{P}=2^{-\dim(C)} \sum_{E\in C} s(E) \overline{E}.
$$
We need therefore understand what happens to an error $E$ when we take
its complex conjugate.  For single qubit errors, this is fairly
straightforward:
$$
\overline{1}=1,\ \overline{\sigma_x}=\sigma_x,\ \
\overline{\sigma_y}=-\sigma_y,\ \overline{\sigma_z}=\sigma_z.
$$
It follows readily that
$$
\overline{E}=(-1)^{\wt_y(E)} E,
$$
where $\wt_y(E)$ is the number of times $\sigma_y$ appears in the tensor
product expansion of $E$.  Now, a fairly straightforward computation gives
us the following identity:
$$
\wt_y(E)\equiv \wt(E)+\langle \sigma_y^{\otimes n}, E\rangle\pmod{2},
$$
where $\sigma_y^{\otimes n}$ is the tensor product of $n$ copies of $\sigma_y$.
Thus
$$
\overline{E}=(-1)^{\wt(E)+\langle \sigma_y^{\otimes n},E\rangle} E.
$$
It follows immediately that an additive code $\Cal{C}$ is real if
and only if the error $\sigma_y^{\otimes n}$ is in $S(C)$.

\proclaim{Theorem 5}
Any additive code is equivalent to a real additive code.
\endproclaim

\demo{Proof}
It suffices to show that any additive code has an element of
weight $n$ in its shadow, since the group of equivalences is transitive
on elements of a given weight.  Now, the number of elements of weight
$n$ is proportional to the coefficient of $y^n$ in $S(x,y)$, or equivalently,
by $S(0,1)$.  But then, by theorem 4, we have:
$$
S(0,1)=A({3\over 2},{1\over 2}).
$$
This is a sum of nonnegative terms, at least one of which is strictly
positive.  Consequently, $S(0,1)>0$, and the theorem is proved.
\qed\enddemo

\head 3. The shadow enumerator for general codes\endhead

The remarks leading up to theorem 5 suggest that a natural starting point
in the generalization of the shadow enumerator involves the conjugate
of $P$.  Consider, therefore, $\Tr(P \overline{P})$.  For an additive
code, this is:
$$
\eqalign{
\Tr(P \overline{P})&=
2^{-2\dim(C)} \sum_{E_1,E_2 \in C} s(E_1)s(E_2)
                     \Tr(E_1 \overline{E_2})\cr
&=
2^{n-2\dim(C)} \sum_{E\in C}
   (-1)^{\wt(E)+\langle\sigma_y^{\otimes n},E\rangle}\cr
&=\cases 2^{n-\dim(C)}&\text{$\sigma_y^{\otimes n}\in S(C)$}\\
         0&\text{otherwise}\endcases\cr}
$$
More generally,
$$
\Tr(P E \overline{P} E)
=
\cases 2^{n-\dim(C)}&\text{$\sigma_y^{\otimes n} E\in S(C)$}\\
       0&\text{otherwise}\endcases
$$
So $E\in S(C)$ if and only if
$$
\Tr(P E \sigma_y^{\otimes n} \overline{P} \sigma_y^{\otimes n} E)=
2^{n-\dim(C)}
$$
Thus the fundamental object seems to be
$$
\tilde{P}=\sigma_y^{\otimes n} \overline{P} \sigma_y^{\otimes n}.
$$

\proclaim{Theorem 6}
Let $M$ be a Hermitian operator on the state space $V$.  Write $M$ as a
linear combination of elements of $\Cal{E}$:
$$
M=\sum_{E\in \Cal{E}} c_E E.
$$
Define $\tilde{M}$ by
$$
\tilde{M}=\sum_{E\in \Cal{E}} (-1)^{\wt(E)} c_E E.
$$
Then
$$
\tilde{M}=\sigma_y^{\otimes n} \overline{M} \sigma_y^{\otimes n}.
$$
Consequently, $\tilde{M}$ is similar to $M$; in particular, if
$M$ is positive semi-definite, then so is $\tilde{M}$.
\endproclaim

\demo{Proof}
Since $M$ is Hermitian, all of the coefficients $c_E$ must be real;
consequently, we may restrict our attention to the case $M=E\in \Cal{E}$.
In that case,
$$
\eqalign{
\tilde{E}&=(-1)^{\wt(E)} E,\cr
         &=(-1)^{\wt_y(E)+\langle \sigma_y^{\otimes n},E\rangle} E\cr
         &=(-1)^{\langle \sigma_y^{\otimes n},E\rangle} \overline{E}\cr
         &=\sigma_y^{\otimes n} \overline{E}\sigma_y^{\otimes n}.\cr}
$$
\qed\enddemo

\proclaim{Corollary 7}
Let $M$ and $N$ be positive semi-definite Hermitian operators on the state
space $V$.  Define
$$
S_d(M,N)=B_d(M,\tilde{N}).
$$
Then for $0\le d\le n$, $S_d(M,N)\ge 0$.
\endproclaim

\demo{Proof}
This follows immediately from theorem 6 and theorem 3.
\qed\enddemo

It remains only to see how $S_d(M,N)$ is related to $A_d(M,N)$.
Define $S(x,y)=\sum_{0\le d\le n} S_d(M,N) x^{n-d} y^d$.  Then

\proclaim{Theorem 8}
$$
S(x,y)=A({x+3y\over 2},{y-x\over 2}).
$$
\endproclaim

\demo{Proof}
Consider the function $W(x,y)$ defined by
$$
W(x,y)=\sum_{0\le d\le n} A_d(M,\tilde{N}) x^{n-d} y^d.
$$
By theorem 1, we have
$$
S(x,y)=W({x+3y\over 2},{x-y\over 2}).
$$
Consequently, it suffices for us to show that $W(x,y)=A(x,-y)$; in
other words, that
$$
A_d(M,\tilde{N})=(-1)^d A_d(M,N).
$$
But
$$
\eqalign{
A_d(M,\tilde{N})&=\sum_{E\in \Cal{E}\atop \wt(E)=d}
                        \Tr(M E)\Tr(\tilde{N} E)\cr
                &=\sum_{E\in \Cal{E}\atop \wt(E)=d}
                        \Tr(M E)(-1)^d \Tr(N E)\cr
                &=(-1)^d A_d(M,N).\cr}
$$
\qed\enddemo

\proclaim{Corollary 9}
For any Hermitian operators $M$, $N$,
$$
\eqalign{
S_d(N,M)&=S_d(M,N)\cr
A_d(\tilde{M},\tilde{N})&=A_d(M,N).\cr}
$$
\endproclaim

\demo{Proof}
The first statement follows immediately from the fact that
$A_d(N,M)=A_d(M,N)$, and the fact that the transform in theorem 8
is independent of $M$ and $N$. The second statement is simply that
$$
S_d(\tilde{N},M)=S_d(M,\tilde{N}),
$$
since $\Tilde{\Tilde{N}}=N$.
\qed\enddemo

This gives us the following theorem (after theorem 21 in \cite\crssII):

\proclaim{Theorem 10 (LP bound for general QECCs)}
If a $((n,K,d))$ exists, then there is a solution to the following
set of linear equations and inequalities:
$$
\eqalign{
&A_0=K^2\cr
&A_i\ge 0\ (0\le i\le n)\cr
&B_i={1\over 2^n} \sum_{0\le r\le n} P_i(r,n) A_r\cr
&A_i=K B_i\ (0\le i<d)\cr
&A_i\le K B_i\ (d\le i\le n)\cr
&S_i={1\over 2^n} \sum_{0\le r\le n} (-1)^r P_i(r,n) A_r\cr
&S_i\ge 0\ (0\le i\le n),\cr}
$$
where
$$
P_i(x,n)=\sum_{0\le s\le i} (-1)^s 3^{i-s} {x\choose s}{n-x\choose i-s}
$$
are the appropriate Krawtchouk polynomials.
\endproclaim

\demo{Proof}
The first five relations come from theorem 1 and 2; the remaining
relations come from theorem 8 and corollary 7.
\qed\enddemo

Remark.  For pure codes, the additional constraint that $A_i=0$ for
$1\le i<d$ must hold.

Using this theorem, one can produce a table of upper bounds analogous
to the table in \cite\crssII.  The resulting table differs in only ten
places:
$$
\hbox to\hsize{
\hfill
\vbox{
\halign{&\hfil $#$ \hfil \quad\cr
((7,2^0)) &((13,2^0)) &((15,2^4)) &((15,2^7)) &((16,2^8))\cr
((18,2^{12})) &((19,2^8)) &((19,2^{13})) &((22,2^{14})) &((25,2^0))\cr}}
\hfill}
$$
In each case, the new bound is precisely 1 greater than the bound for
additive codes.  Consequently, nearly all of the codes in \cite\crssII\ that
are optimal among additive codes are optimal among all codes; in
particular, for $1\le n\le 12$, the only place where the bound is not known
to be tight is $n=7,k=0$.  It is also worth noting that, just as for
additive codes, the LP bound for impure codes agrees with the LP bound for
pure codes for all $n$ checked ($1\le n\le 30$).

\head 4. Parity issues; self-dual codes\endhead 

In the study of additive codes, one important distinction is between
even codes (those that contain no element of odd weight) and
odd codes (those in which half of the elements have odd weight).
This distinction carries over to general codes, using shadow theory.

\proclaim{Definition}
A code $\Cal{C}$ with projection matrix $P$ is even if $P=\tilde{P}$, and
odd if $\Tr(P \tilde{P})=0$.
\endproclaim

Remark.  Note that the typical nonadditive code is neither even nor
odd.

If $\Cal{C}$ is odd, we can define a new code $\Cal{C}_0$,
called the ``even subcode'', as the image of the projection
$$
P+\tilde{P}.
$$
(Note that the even subcode of a code is actually {\it larger};
the terminology is by analogy with the additive case.)

\proclaim{Theorem 11}  Let $\Cal{C}$ be an odd code, and let $\Cal{C}_0$
be its even subcode.  Then
$$
\eqalign{
A_d(\Cal{C}_0)&=\cases
 4 A_d(\Cal{C})& \text{$d\equiv 0\pmod 2$}\\
 0&\text{$d\equiv 1\pmod 2$}\endcases\cr
B_d(\Cal{C}_0)&=2(B_d(\Cal{C})+S_d(\Cal{C})).\cr}
$$
\endproclaim

\demo{Proof}
First $A_d$:
$$
\eqalign{
A_d(\Cal{C}_0)&=A_d(P+\tilde{P})\cr
               &=A_d(P)+A_d(\tilde{P})+2 A_d(P,\tilde{P})\cr}
$$
Since $A_d(P,\tilde{P})=(-1)^d A_d(P)$, the result follows immediately.
Similarly,
$$
\eqalign{
B_d(\Cal{C}_0)&=B_d(P)+B_d(\tilde{P})+2 B_d(P,\tilde{P})\cr
               &=2(B_d(P)+S_d(P)).\cr}
$$
\qed\enddemo

\smallskip

An interesting thing happens with the shadow enumerator for self-dual
codes (that is, codes with $K=1$).  In this case, $P$ has rank 1,
so may be written as $v v^\dagger$, with $v$ a unit vector.
In this case, we have
$$
\eqalign{
S_d(P)&=
\sum_{E\in \Cal{E}\atop \wt(E)=d}
Tr(v v^\dagger E \sigma_y^{\otimes n} \overline{v} v^t
                 \sigma_y^{\otimes n})\cr
&=
\sum_{E\in \Cal{E}\atop \wt(E)=d}
|v^t E \sigma_y^{\otimes n} v|^2.\cr}
$$
Now,
$$
\eqalign{
(E\sigma_y^{\otimes n})^t
&=
(-1)^n \sigma_y^{\otimes n} E^t\cr
&=
(-1)^n \sigma_y^{\otimes n} \overline{E}\cr
&=
(-1)^{n+\langle \sigma_y^{\otimes n},E\rangle} \overline{E}
\sigma_y^{\otimes n}\cr
&=
(-1)^{n-\wt(E)} E \sigma_y^{\otimes n}.\cr}
$$
In particular, if $n-\wt(E)$ is odd, then $E \sigma_y^{\otimes n}$
is antisymmetric, and $\Tr(P E \tilde{P} E)=0$.  Consequently:

\proclaim{Theorem 12}
Let $\Cal{C}$ be a self-dual quantum code.  Then
$$
S_{n-2k-1}(\Cal{C})=0
$$
for $0\le k\le \lfloor {n-1\over 2}\rfloor$.
\endproclaim

\proclaim{Corollary 13}
A self-dual quantum code is odd whenever $n$ is odd.
\endproclaim

\demo{Proof}
Consider $S_0(\Cal{C})$.
\qed\enddemo

We can now state the following result:

\proclaim{Theorem 14}
If a (pure) $((6m+l,1,d))$ exists, with $0\le l\le 5$,
then
$$
d\le\cases 2m+2&\text{$l<5$}\\
           2m+3&\text{$l=5$}\endcases
$$
If a $((6m+5,1,2m+3))$ exists (necessarily odd), then so does
a $((6m+6,1,2m+4))$.  Finally, any $((6m,1,2m+2))$ must be even.
\endproclaim

\demo{Proof}
The proof is outside the scope of this paper; see \cite\shadowboundcodes\
for more details.  (It should be noted that the version in
\cite\shadowboundcodes\ is stated in terms of additive codes; however, the
proof makes no assumptions of integrality, so carries over directly.)

The only thing remaining is to give the construction of a
$((6m+6,1,2m+4))$ from a $((6m+5,1,2m+3))$.  Let $\Cal{C}$ be
a $((6m+5,1,2m+3))$.  From the proof in \cite\shadowboundcodes, we have
$S_i(\Cal{C})=0$ for $0\le i<2m+3$.  Now, taking the even subcode
of $\Cal{C}$ gives us a code with $A_i(\Cal{C}_0)=0$ for
$0\le i<2m+4$, and $B_i(\Cal{C}_0)=0$ for $0\le i<2m+3$.
Now \cite\unitaryenum gives a construction that produces a new self-dual code
$\Cal{C}'$ of length $6m+6$ with
$$
A_i(\Cal{C}')={1\over 4} (A_i(\Cal{C}_0)-A_{i-1}(\Cal{C}_0))
             +{1\over 2} B_{i-1}(\Cal{C}_0)
$$
But then $A_i(\Cal{C}')=0$ for $0\le i<2m+4$, and $\Cal{C}'$ is
the desired $((6m+6,1,2m+4))$.
\qed\enddemo

Also from \cite\shadowboundcodes, we get the following:

\proclaim{Theorem 15}
If a $((6m-1+l,K,d))$ exists for $K>1$, with $0\le l\le 5$, then
$$
d\le\cases 2m+1	&\text{$l<5$}\\
           2m+2 &\text{$l=5$}\endcases
$$
Moreover, any $((6m-1,K,2m+1))$ is the even subcode of a $((6m-1,1,2m+1))$
(in particular, $K=2$).
\endproclaim

In particular, any quantum code of length $n$ can correct at most
$\lfloor{n+1\over 6}\rfloor$ errors.

\head Conclusion\endhead

We have extended the work of Shor and Laflamme by defining another
nonnegative enumerator, computable in terms of their enumerators.
This further strengthens their linear programming bound, to the
point that the best bounds for general codes are nearly the same
as the best bounds for additive codes.  We also extended a bound
on additive codes proved using shadow theory to general codes,
obtaining as a consequence that any code of length $n$ can
correct at most $\lfloor{n+1\over 6}\rfloor$ errors.

\head Acknowledgements\endhead

We would like to thank P. Shor and N. Sloane for many helpful discussions.

\enddocument
\bye